\documentclass[aps,prl,showpacs,superscriptaddress]{revtex4}

\usepackage{graphics}% Include figure files
\usepackage{graphicx}% Include figure files
\usepackage{dcolumn}% Align table columns on decimal point
\usepackage{bm}% bold math
\usepackage{color}
\usepackage{amssymb,amsmath}

\newcommand{\de}{\partial}

\newcommand{\eps}{\epsilon}

\begin{document}

\title{Soliton self-frequency blue-shift in gas-filled hollow-core photonic crystal fibers}
\author{Mohammed F. Saleh}
\author{Wonkeun Chang}
\author{Philipp H\"olzer}
\author{Alexander Nazarkin}
\author{John C. Travers}
\affiliation{Max Planck Institute for the Science of Light, G\"{u}nther-Scharowsky str. 1, 91058 Erlangen, Germany}

\author{Nicolas Y. Joly}
\author{Philip St.J. Russell}
\affiliation{Max Planck Institute for the Science of Light, G\"{u}nther-Scharowsky str. 1, 91058 Erlangen, Germany}
\affiliation{Department of Physics, University of Erlangen-Nuremberg, Germany}

\author{Fabio Biancalana}
\affiliation{Max Planck Institute for the Science of Light, G\"{u}nther-Scharowsky str. 1, 91058 Erlangen, Germany}
\date{\today}

\begin{abstract}
We show theoretically that the photoionization process in a hollow-core photonic crystal fiber filled with a Raman-inactive noble gas leads to a constant acceleration of solitons in the time domain with a continuous shift to higher frequencies, limited only by ionization loss. This phenomenon is opposite to the well-known Raman self-frequency red-shift of solitons in solid-core glass fibers. We also predict the existence of unconventional long-range non-local soliton interactions leading to spectral and temporal soliton clustering. Furthermore, if the core is filled with a Raman-active molecular gas, spectral transformations between red-shifted, blue-shifted and stabilized solitons can take place in the same fiber.
\end{abstract}
\maketitle

\begin{figure}
\includegraphics[width=8cm]{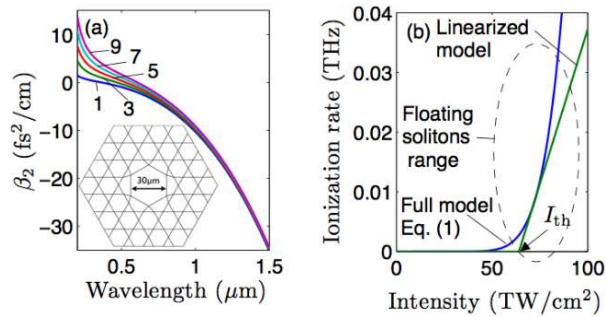}
\caption{(Color online). (a) GVD of an Ar-filled HC-PCF for gas pressures between 1 and 9 bar (calculated from \cite{Marcatili64}). All subsequent calculations in this paper assume 5 bar pressure. Inset: cross-section of a broadband-guiding HC-PCF with a kagom\'{e}-lattice cladding and a core diameter $30$ $\mu$m. Typical experimental transmission losses for the fundamental mode are 1 dB/m for at $800$ nm. (b) Comparison of the dependence of the Ar ionization rate on the pulse intensity using the full model of Eq. (\ref{eqW}) and the linearized model.  \label{fig1}}
\end{figure}

\paragraph{Introduction ---}
Hollow-core photonic crystal fibers (HC-PCFs) \cite{Russell03} based on a kagom\'{e}-lattice claddings have recently been shown to be very interesting for the investigation of broadband light-matter interactions between intense optical pulses and gaseous media.  The fibers typically show transmission bands covering the visible and near-IR parts of the spectrum with relatively low loss and low group velocity dispersion (GVD), absence of surface modes, and high confinement of light in the core. Filled with a noble gas, they have recently been used in high-harmonic and efficient deep UV generation from fs pump pulses at 800 nm \cite{Heckl09,Joly11}. It has been previously shown that the Raman threshold can be drastically reduced in a HC-PCF filled with a Raman-active gas (such as H$  _{2}$) \cite{Benabid02a}. The system can be used for detailed experimental studies of, e.g., self-similar solutions of the sine-Gordon equation \cite{Nazarkin10} and backward stimulated Raman scattering \cite{Benabid02a,Abdolvand09}. In bandgap-guiding gas-filled HC-PCFs, which have much narrower bands of transmission, a limited ionization-induced blue-shift of guided ultrashort pulses has been reported \cite{Serebryannikov07, Fedotov07}. Very recently, ultrafast nonlinear dynamics in the ionization regime has been studied experimentally in Ar-filled kagom\'{e}-style HC-PCF \cite{Hoelzer11a} (a detailed account of these experiments is available in a parallel submission \cite{Hoelzer11b}). The reasons for the success of kagom\'{e} HC-PCF in these applications are: (i) a group velocity dispersion (GVD) that is remarkably small ($ |\beta_{2}|<10 $ fs$^{2}$/cm $ \equiv 1 $ ps$^{2}$/km  from 400 to 1000 nm) in comparison to solid-core fibers (Fig. \ref{fig1}(a)) \cite{Joly11,Hoelzer11b}; (ii) the gas and waveguide contributions to the GVD can be balanced by varying the pressure, unlike in large-bore capillary-based systems where the normal dispersion of the gas dominates over the waveguide dispersion \cite{Nold10}.

Photoionization in gases is traditionally modeled using the full electric field of the pulse \cite{Geissler99}. In this paper, we first develop a new model to study pulse propagation in gas-filled HC-PCFs in terms of the complex {\em envelope} of the pulse. Using this model, we show analytically for the first time that intra-pulse photoionization leads to: (i) a soliton self-frequency blue-shift; (ii) long-range ``non-local" soliton correlations and clustering; and (iii) spectral transformations of red-shifted, blue-shifted and stabilized solitons in Raman-active gas-filled HC-PCFs.

\paragraph{Governing equations ---}
Photoionization can take place by either tunneling or multiphoton processes. These regimes are characterized by the Keldysh parameter $p_{\rm K}$ \cite{Geissler99,Wegener05}. In the tunneling regime ($p_{\rm K}\ll 1$) the time-averaged ionization rate $\mathcal{W}(I)$ is given by  \cite{Keldysh64,Sprangle02}
\begin{equation}
\mathcal{W}(I)=d\,(I_{H}/I) ^{1/4}\, \exp[ -b\,(I_{H}/I)^{1/2}],\label{eqW}
\end{equation}
where $ d\equiv 4\,\Omega_{0}\,[3/\pi]^{1/2}\,[U_{I}/U_{H}]^{7/4} $, $b\equiv 2/3\, [U_{I}/U_{H}]^{3/2} $, $ \Omega_{0}=4.1\times 10^{16}$ Hz is the characteristic atomic frequency, $ U_{I} $ is the ionization energy of the gas ($\sim 15.76$ eV for argon), $ U_{H}\approx 13.6 $ eV is the ionization energy of hydrogen, $ I_{H} =3.6 \times 10^{16} $ W/cm$^{2}$ and $I$ is the laser pulse intensity. For values of $I$ in the range of $ 100 $ TW/cm$ ^{2} $, the Keldysh parameter is $p_{\rm K}\lessapprox 1$ for noble gases. However, experiments show that tunneling models provide excellent agreement with the experimental measurements even for $p_{\rm K}\approx1$ \cite{Gibson90,Augst91}. As shown in Fig. \ref{fig1}(b), Eq. (\ref{eqW})predicts an ionization rate that is exponential-like for pulse intensities above a threshold value. Loss due to absorption of photons in the plasma is \textit{proportional} to the ionization rate. Hence, any pulse with $ I\gg I_{\mathrm{th}} $ will have its intensity strongly driven back to near the threshold value, resulting in drastically reduced ionization loss.

This allows us to use the first-order Taylor series to \textit{linearize} the tunneling model just above $ I= I_{\mathrm{th}} $, where the optical pulses can survive for relatively long time without appreciable attenuation. Expanding Eq. (\ref{eqW}) in its linear regime around an arbitrary point $a=I_{a}/I_{H} $, results in $\mathcal{W}\approx \tilde{\sigma}\, \Delta I\, \Theta \left( \Delta I \right)$, where $ \Delta I \equiv I -I_{\mathrm{th}} $, $ \tilde{\sigma}=d\,  e^{-x}\,(2x-1)/[4\, a^{5/4}\,I_{H}] $, $ I_{\mathrm{th}}=a\,I_{H} (2x-5)/(2x-1) $ is the threshold intensity, $ x=b/\sqrt{a} $, and $a$ is chosen to reproduce the physically observed threshold intensity in the fiber of Fig. \ref{fig1}(a), $a\cong 2\times 10^{-3}$. The purpose of the Heaviside function $ \Theta $ is to set the ionization rate to zero below the threshold intensity, see Fig. \ref{fig1}(b).

One can prove from first principles that propagation of light in a HC-PCF filled with an ionized Raman-active gas can be then described by the following coupled equations:
\begin{equation}
\begin{array}{l}
\left[ i\de_{z}+\hat{D}(i\de_{t})+\gamma_{\rm K} R(t)\otimes|\Psi(t)|^{2} -\dfrac{\omega_{\rm p}^{2}}{2k_{0}c^{2}}+i\alpha\right] \Psi  =0  \\
\de_{t}n_{\rm e}=[\tilde{\sigma}/A_{\mathrm{eff}}]\left[n_{\rm T}-n_{\rm e}\right] \Delta|\Psi|^{2}\, \Theta \left(\Delta|\Psi|^{2} \right)
\end{array} , \label{eq1}
\end{equation}
where $\Psi(z,t)$ is the electric field {\em envelope}, $z$ the longitudinal coordinate along the fiber, $t$ is the time in a reference frame moving with the pulse group velocity, $\hat{D}(i\de_{t})\equiv\sum_{m\geq 2}\beta_{m}(i\de_{t})^{m}/m!$ is the full dispersion operator, $\beta_{m}$ is the $m$-th order dispersion coefficient calculated at an arbitrary reference frequency $\omega_{0}$, $\gamma_{\rm K}$ is the Kerr nonlinear coefficient of the gas, $R(t)= (1-\rho)\delta(t)+\rho\, h(t)$ is the normalized Kerr and Raman response function of the gas, $\delta(t)$ is the Dirac delta function, $\rho$ is the relative strength of the non-instantaneous Raman nonlinearity, $h(t)$ is the causal Raman response function of the gas \cite{Serebryannikov07,Agrawal07}, the symbol $\otimes$ denotes the time convolution, $c$ is the speed of light, $k_{0}=\omega_{0}/c$, $ \omega_{0} $ is the pulse central frequency, $\omega_{\rm p}=[e^{2} n_{\rm e}/(\eps_{0}m_{\rm e})]^{1/2}$ is the plasma frequency associated with an electron density $n_{\rm e}(t)$, $e$ and $m_{\rm e}$ are the electron charge and mass, and $\eps_{0}$ is the vacuum permittivity, $ \alpha=\alpha_{1}+\alpha_{2} $ is the total loss coefficient, $ \alpha_{1} $ is the fiber loss, $ \alpha_{2}=\frac{A_{\mathrm{eff}}U_{I}}{2|\Psi|^{2}}\,\de_{t}n_{\rm e} $ is the ionization-induced loss term, $A_{\mathrm{eff}} $ is the effective mode area, $ \Delta|\Psi|^{2}= |\Psi|^{2} -|\Psi|^{2}_{\mathrm{th}}$, $ |\Psi|^{2}=I A_{\mathrm{eff}} $, $ |\Psi|_{\mathrm{th}}^{2}=I_{\mathrm{th}} A_{\mathrm{eff}} $, and $n_{\rm T}$ is the total number density of ionizable atoms in the fiber, associated with the maximum plasma frequency $\omega_{\rm T}\equiv[e^{2} n_{\rm T}/(\eps_{0}m_{\rm e})]^{1/2}$. In these coupled equations, the recombination process is neglected since the pulse duration (of the order of tens of fs) is always shorter than the recombination time \cite{Wood93}. If $|\Psi|^{2}$ is measured in W, $\tilde{\sigma}/c\,A_{\mathrm{eff}}\equiv\gamma_{\rm I}$ has the dimensions of W$^{-1}$m$^{-1}$. This is the nonlinearity associated with the plasma formation in the fiber. According to recent experimental measurements \cite{Borzsonyi10}, $\gamma_{\rm K}$ shows a linear dependence on the gas pressure. These coupled equations (\ref{eq1}) are the first contribution of this paper. The validity of Eqs. (\ref{eq1}) has been verified using a more complete ionization  model based on the unidirectional wave equation \cite{Chang11}.

\paragraph{Perturbation theory for floating pulses ---} In order to extract useful analytical information from Eqs. (\ref{eq1}), further simplifications are necessary. For pulses with maximum intensities just above the ionization threshold (which we dub {\em floating} pulses, a new concept introduced in this paper for the first time), the ionization loss is not large and can be neglected as a first approximation. For such pulses, only a small portion of energy above the threshold intensity contributes to the creation of free electrons. Furthermore, for floating pulses one can remove the $\Theta$-function from the equations, provided that the cross-section $\tilde{\sigma}$ is replaced by a properly reduced $\tilde{\sigma}'$ that takes into account the overestimation of the ionization rate \cite{note1}. Introducing the following rescalings and redefinitions: $\xi\equiv z/z_{0}$, $\tau\equiv t/t_{0}$, $\Psi_{0}\equiv[\gamma_{\rm K}z_{0}]^{-1/2}$, $\psi\equiv \Psi/\Psi_{0}$, $r(\tau)\equiv R(t)\,t_{0}$, $\phi\equiv \frac{1}{2}k_{0}z_{0}\,[\omega_{\rm p}/\omega_{0}]^{2}$, $\phi_{\rm T}\equiv \frac{1}{2}k_{0}z_{0}\,[\omega_{\rm T}/\omega_{0}]^{2}$, and $\sigma\equiv\tilde{\sigma}'\,t_{0}/[ A_{\mathrm{eff}}\gamma_{\rm K}\,z_{0}]$, where $z_{0}\equiv t_{0}^{2}/|\beta_{2}(\omega_{0})|$ is the second-order dispersion length at the reference frequency $\omega_{0}$ and $t_{0}$ is the input pulse duration \cite{Agrawal07}. Hence, the two coupled equations for floating pulses can be replaced by
\begin{equation}
\begin{array}{l}
\left[ i\de_{\xi}+\hat{D}(i\de_{\tau})+r(\tau)\otimes|\psi(\tau)|^{2}-\phi\right]\psi=0 \\
\de_{\tau}\phi=\sigma(\phi_{\rm T}-\phi) |\psi|^{2}
\end{array}.\label{eq2}
\end{equation}
The total number of photons is conserved in this set of coupled equations -- in contrast to Eqs. (\ref{eq1}) -- since losses are neglected for floating pulses.

The effect of the Raman and ionization perturbations on the soliton dynamics in HC-PCFs can be studied using Eqs. (\ref{eq2}). The second equation can be solved analytically, $ \phi(\tau)=\phi_{\rm T}\left\lbrace 1 - \exp\left[-\sigma\int_{-\infty}^{\tau}|\psi(\tau')|^{2}d\tau'\right]\right\rbrace $, with the initial condition $\phi(-\infty)=0$, corresponding to the absence of any plasma before the pulse arrives. For a small ionization cross-section, $ \phi(\tau)\simeq \eta\int_{-\infty}^{\tau}|\psi(\tau')|^{2}d\tau' $, where $\eta\equiv\sigma\phi_{\rm T}$. Moreover, in the long-pulse limit $|\psi(\tau-\tau')|^{2}\simeq |\psi(\tau)|^{2}-\tau'\de_{\tau}|\psi(\tau)|^{2}$  \cite{Agrawal07}. This allows the two coupled equations to be reduced to a single partial integro-differential equation:
\begin{equation}
i\de_{\xi}\psi+\hat{D}(i\de_{\tau})\psi+|\psi|^{2}\psi-\tau_{\rm R}\psi\de_{\tau}|\psi|^{2}-\eta\psi\int_{-\infty}^{\tau}\!\!\!\!\!\!\!\!|\psi|^{2}d\tau' =0
\label{finalx}
\end{equation}
where $\tau_{\rm R}\equiv\int_{0}^{\infty} \tau'\,r(\tau')\, d\tau'$. This equation shows clearly that the effect of ionization is exactly opposite to that of the Raman effect: the fourth term in Eq. (\ref{finalx}) involves a {\em derivative} of the field intensity, while the fifth term involves an {\em integral} on the same quantity. One can then conjecture that the last term will lead to a {\rm soliton self-frequency blue-shift} due to ionization, instead of a red-shift. To prove this statement, we use the perturbation theory described in \cite{Agrawal07}. First, the soliton functional shape is assumed to be unchanged during the action of the perturbations induced by the Raman effect and the photoionization process (this must be verified {\em a posteriori}): $\psi_{\rm S}(\xi,\tau)=A_{0}{\rm sech}\left[A_{0}(\tau-\tau_{\rm p}(\xi))\right]e^{-i\Omega(\xi)\tau}$, with $\tau_{\rm p}(\xi)$ is the temporal location of the soliton peak and $\Omega(\xi)$ is the self-frequency shift. When this {\em Ansatz} is inserted into Eq. (\ref{finalx}), simple ordinary differential equations can be obtained for both $\Omega(\xi)$ and $\tau_{\rm p}(\xi)$, results in $\Omega(\xi)=\Omega_{\rm Raman}(\xi)+\Omega_{\rm ion}(\xi)=-g\,\xi  $, $\tau_{\rm p}(\xi)=g\,\xi^{2}/2$, and $ g= g_{\rm red}+g_{\rm blue}$, where $g_{\rm red}=+(8/15)\tau_{\rm R}A_{0}^{4}$ and $g_{\rm blue}=-(2/3)\eta A_{0}^{2}$. Note that $g$ can be positive, negative or even zero, depending on the value of $\eta$, $\tau_{\rm R}$ and $A_{0}$. By using the exact solution for $\phi(\tau)$ given previously, one obtains the more precise rate
$g'_{\rm blue}=\sigma^{-2}A_{0}^{-1}\phi_{\rm T}\left[(1-\sigma A_{0})-(1+\sigma A_{0})\exp(-2\sigma A_{0})\right]$, which tends to $g_{\rm blue}$ for small values of $\sigma$, but starts to differ considerably from it for $A_{0}>\sigma^{-1}$. The above solution clearly shows that, in the range of validity of perturbation theory (i.e., for floating solitons), {\em photoionization leads to a soliton self-frequency blue-shift}. This blue-shift is accompanied by a constant acceleration of the pulse in the time domain -- opposite to the Raman effect, which produces pulse deceleration. The blue-shift is limited only by ionization loss, which slowly decreases the pulse intensity until it falls below the threshold value. Other important effects, such as the formation of inverted gravity-like bound solitons supported by the plasma nonlinearity, will be reported elsewhere.

\begin{figure}
\includegraphics[width=14cm]{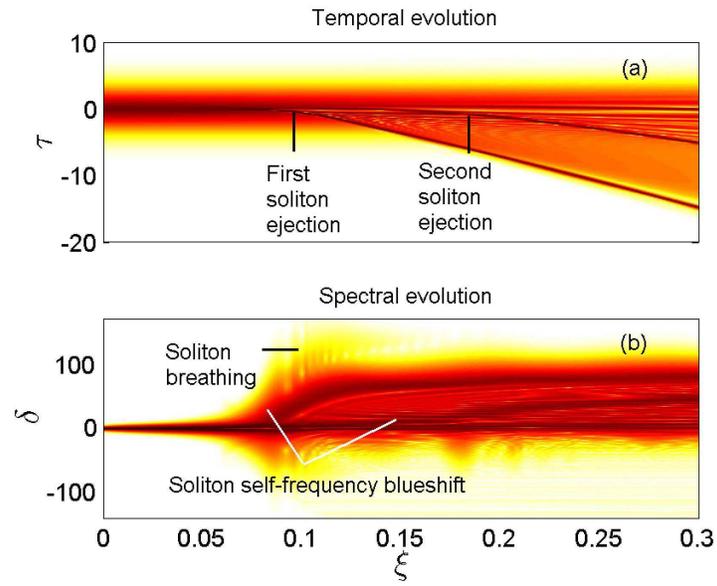}
\caption{(Color online). Temporal (a) and spectral (b) evolution of an energetic pulse propagating in the Ar-filled HC-PCF. The temporal profile of the input pulse is $ N\,\mathrm{sech}\,\tau $, with $ N=8 $. The panels show the ejection of two fundamental solitons that continuously blue-shift until ionization loss reduces their intensities below the threshold value.
\label{fig2}}
\end{figure}

In the presence of ionization-induced losses above the threshold intensity, Eqs. (\ref{eq1}) must be numerically solved to study the full dynamics of floating pulses. Figures \ref{fig2}(a,b) show the temporal and spectral evolution of a high-order input soliton, closely following the results reported in the companion experimental paper \cite{Hoelzer11b}. When the intensity of the energetic pulse exceeds the threshold value as a result of self-compression, a fundamental soliton is ejected from the main pulse and continues to blue-shift until ionization loss reduces its amplitude below the threshold value. At longer distances, another compression occurs and a second soliton is generated.  The use of a kagom\'{e}-style HC-PCF is essential to observe the soliton blue-shift, since conventional photonic-bandgap fibers have much stronger dispersion variations, which would quickly destabilize any possible solitary wave as in \cite{Fedotov07,Serebryannikov07}.

\paragraph{Long-range non-local soliton forces and clustering ---} An interesting and unexpected interaction occurs between two solitons when their temporal separation is shorter than the recombination time, due to the non-vanishing electron density tail. Using the exact formula for the ionization field $\phi(t)$, one can see that a leading soliton with amplitude $ A_{0} $ can slow down the acceleration of a trailing soliton by an exponential factor $\exp(-2\sigma A_{0})$. The reason is that the ionization field $\phi(t)$, created by the first soliton, decays at a relatively slow rate. This establishes a unique non-local interaction between this soliton and other temporally distant solitons.

\begin{figure}
\includegraphics[width=14cm]{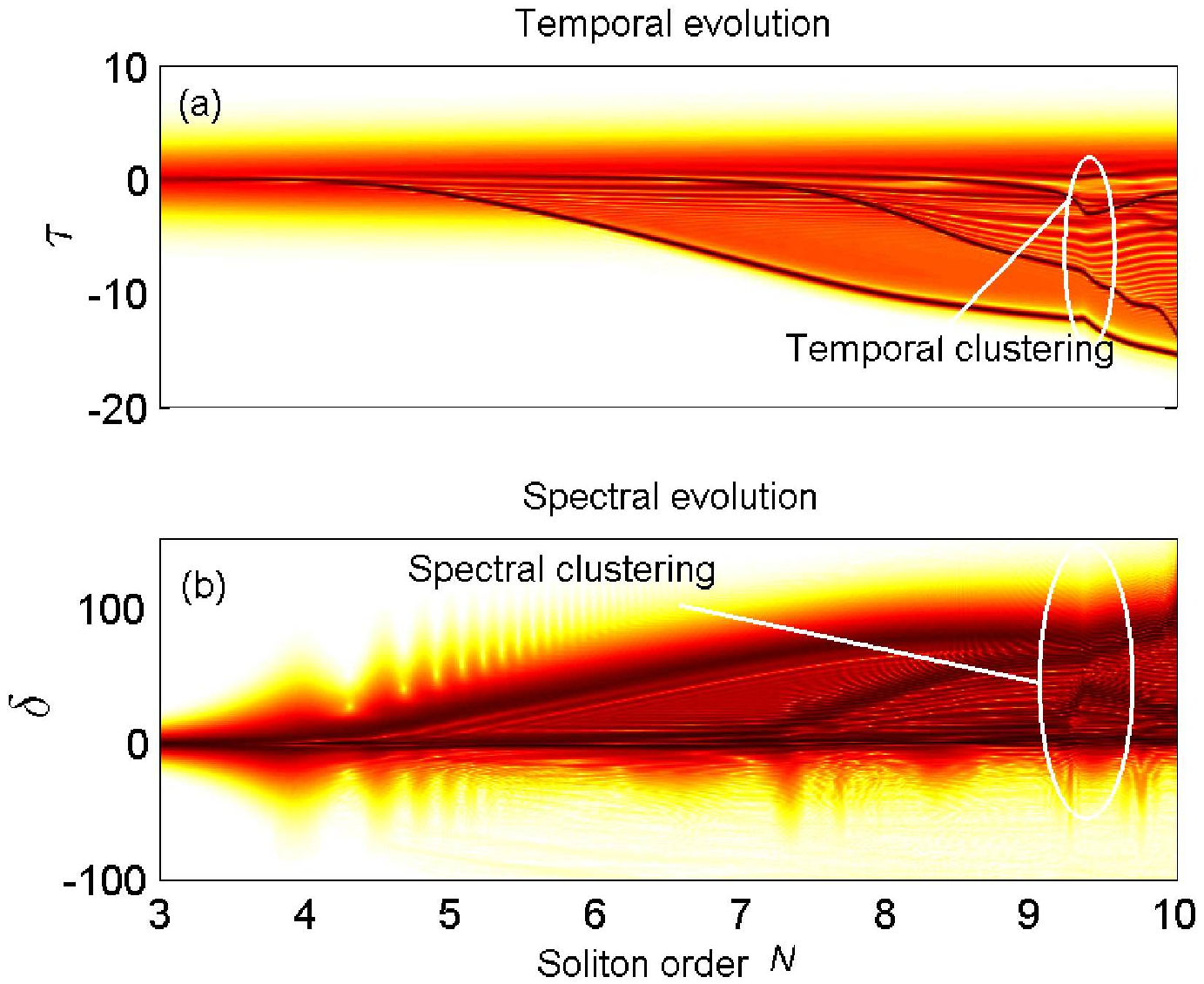}
\caption{(Color online). Temporal (a) and spectral (b) outputs of an energetic pulse $ N\,\mathrm{sech}\,\tau $ after propagating inside an Ar-filled HC-PCF with length $\xi= 1/4 $ versus the soliton order $ N $. Temporal and spectral clustering occur at $ N=9.2 $ due to the long-range ``non-local" soliton interactions described in the text.
\label{fig3}}
\end{figure}

Figures \ref{fig3}(a,b) show the output temporal and spectral dependence of a pulse $ N\,\mathrm{sech}\,\tau $ on the soliton order $ N $.   In the presence of ionization loss, when the intensity of the leading soliton decreases to the threshold value, i.e., the blue-shifting process ceases, the lagging soliton will recover its expected blue-shift. The reason is the disappearance of the exponential decaying factor at that particular point. Also, there is a maximum frequency attained by each soliton that depends on the initial soliton intensity. These interactions may lead, at some `magic' input energy, to clustering two or more distinct solitons in both temporal and spectral domains, as shown in Figs. \ref{fig3}(a,b).

%This factor may lead to a swapping of the positions of the generated fundamental solitons from the fission of an energetic input pulse.

\paragraph{Soliton spectral transformations ---}
Interestingly, this perturbation theory for floating solitons predicts the formation of spectrally stabilized solitons in Raman-active gases due to the different signs and $A_{0}$-dependence of the Raman and photoionization shifts, $g_{\rm red}\propto A_{0}^{4}$ (a well-known result \cite{Agrawal07}), and $g_{\rm blue}\propto -A_{0}^{2}$ (reported for the first time in this paper). If one launches an sufficiently energetic pulse into the fiber, soliton fission takes place, independent of the particular perturbation applied \cite{Lucek92,Husakou01}. This generates a train of fundamental solitons with progressively decreasing peak amplitudes.

\begin{figure}
\includegraphics[width=14cm]{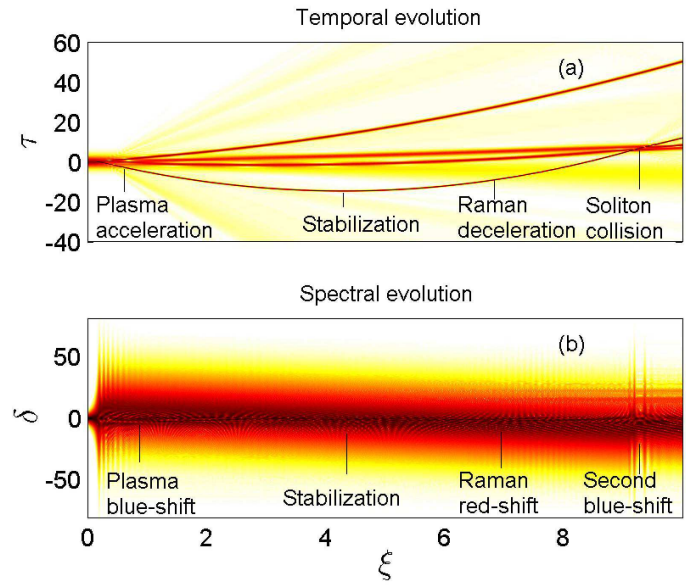}
\caption{(Color online). Temporal (a) and spectral (b) evolution of an energetic pulse propagates in a HC-PCF filled with argon and air. The temporal profile of the input pulse is $ N\,\mathrm{sech}\,\tau $, with $ N=4 $. Soliton temporal and spectral trajectories show an initial acceleration and blue-shift due to plasma formation, an intermediate stabilization below-threshold intensity, and finally a deceleration and red-shift due to Raman effect. Near the fiber end a second blue-shift event takes place due to soliton collision, generating a second surge of ionized plasma.
\label{fig4}}
\end{figure}

The temporal and spectral evolution of such a pulse when it propagates in a mixture of argon and air (Raman-active) is depicted in Figs. \ref{fig4}(a,b). After the fission process, solitons with intensities less than the threshold intensity are red-shifted by the undisturbed Raman process. However, solitons possessing intensities above the threshold value are influenced simultaneously by both the photoionization and the Raman effects. Depending on their initial intensities, these solitons can be \textit{initially} blue-shifted, red-shifted or stabilized. The Raman self-frequency red-shift will be more pronounced than the ionization self-frequency blue-shift for floating solitons possessing initially larger amplitudes ($A_{j}>A_{\rm cr}$), where $A_{\rm cr}$ is a critical amplitude. However, for less intense floating solitons ($A_{j}<A_{\rm cr}$) it may happen that exactly the opposite phenomenon occurs, i.e., the blueshift will dominate. The critical intensity can be estimated from the equation $g_{\rm red}+g_{\rm blue}=0$, giving $A_{\rm cr}^{2}=5\eta/(4\tau_{\rm R})$. When the ionization loss arrests the photoionization process, the initially blue-shifted solitons start to reverse their self-frequency shift towards the red. At a certain point, these solitons can become frequency-stabilized over a short distance. This may result in multiple collisions between floating solitons if their temporal trajectories intersect. When the instantaneous intensity exceeds the threshold value upon collision, a second blue-shift event may occur.

\paragraph{Conclusions ---} A direct photoionization process can act on solitons by constantly blue-shifting their central frequencies  representing the exact counterpart of the Raman self-frequency red-shift  when the intensity of solitons is slightly above the photo-ionization threshold. This spectral transformation is limited by the ionization loss that restricts the pulse intensity to the threshold value, hence arresting the soliton blue-shift. The new theoretical model, presented by Eqs. (\ref{eq1}), is suitable for analytical manipulations, and has led us to predict a number of new phenomena such as long-range non-local correlation forces, and spectral transformation between red- and blue-shift in Raman-active gases. The results reveal new physics and offer novel opportunities for the manipulation and control of the soliton dynamics inside these versatile optical waveguides.

\bibliographystyle{apsrev4-1}	
%\bibliography{Refs_paper1}	

\begin{thebibliography}{26}%
\makeatletter
\providecommand \@ifxundefined [1]{%
 \@ifx{#1\undefined}
}%
\providecommand \@ifnum [1]{%
 \ifnum #1\expandafter \@firstoftwo
 \else \expandafter \@secondoftwo
 \fi
}%
\providecommand \@ifx [1]{%
 \ifx #1\expandafter \@firstoftwo
 \else \expandafter \@secondoftwo
 \fi
}%
\providecommand \natexlab [1]{#1}%
\providecommand \enquote  [1]{``#1''}%
\providecommand \bibnamefont  [1]{#1}%
\providecommand \bibfnamefont [1]{#1}%
\providecommand \citenamefont [1]{#1}%
\providecommand \href@noop [0]{\@secondoftwo}%
\providecommand \href [0]{\begingroup \@sanitize@url \@href}%
\providecommand \@href[1]{\@@startlink{#1}\@@href}%
\providecommand \@@href[1]{\endgroup#1\@@endlink}%
\providecommand \@sanitize@url [0]{\catcode `\\12\catcode `\$12\catcode
  `\&12\catcode `\#12\catcode `\^12\catcode `\_12\catcode `\%12\relax}%
\providecommand \@@startlink[1]{}%
\providecommand \@@endlink[0]{}%
\providecommand \url  [0]{\begingroup\@sanitize@url \@url }%
\providecommand \@url [1]{\endgroup\@href {#1}{\urlprefix }}%
\providecommand \urlprefix  [0]{URL }%
\providecommand \Eprint [0]{\href }%
\providecommand \doibase [0]{http://dx.doi.org/}%
\providecommand \selectlanguage [0]{\@gobble}%
\providecommand \bibinfo  [0]{\@secondoftwo}%
\providecommand \bibfield  [0]{\@secondoftwo}%
\providecommand \translation [1]{[#1]}%
\providecommand \BibitemOpen [0]{}%
\providecommand \bibitemStop [0]{}%
\providecommand \bibitemNoStop [0]{.\EOS\space}%
\providecommand \EOS [0]{\spacefactor3000\relax}%
\providecommand \BibitemShut  [1]{\csname bibitem#1\endcsname}%
\let\auto@bib@innerbib\@empty
%</preamble>
\bibitem [{\citenamefont {{P.~{St.J}.~Russell}}(2003)}]{Russell03}%
  \BibitemOpen
  \bibfield  {author} {\bibinfo {author} {\bibnamefont
  {{P.~{St.J}.~Russell}}},\ }\href@noop {} {\bibfield  {journal} {\bibinfo
  {journal} {Science}\ }\textbf {\bibinfo {volume} {299}},\ \bibinfo {pages}
  {358} (\bibinfo {year} {2003})}\BibitemShut {NoStop}%
\bibitem [{\citenamefont {Heckl}\ \emph {et~al.}(2009)\citenamefont {Heckl},
  \citenamefont {Baer}, \citenamefont {Kr\"{a}nkel}, \citenamefont {Marchese},
  \citenamefont {Schapper}, \citenamefont {Holler}, \citenamefont
  {S\"{u}dmeyer}, \citenamefont {Robinson}, \citenamefont {Tisch},
  \citenamefont {Couny}, \citenamefont {Light}, \citenamefont {Benabid},\ and\
  \citenamefont {Keller}}]{Heckl09}%
  \BibitemOpen
  \bibfield  {author} {\bibinfo {author} {\bibfnamefont {O.~H.}\ \bibnamefont
  {Heckl}}, \bibinfo {author} {\bibfnamefont {C.~R.~E.}\ \bibnamefont {Baer}},
  \bibinfo {author} {\bibfnamefont {C.}~\bibnamefont {Kr\"{a}nkel}}, \bibinfo
  {author} {\bibfnamefont {S.~V.}\ \bibnamefont {Marchese}}, \bibinfo {author}
  {\bibfnamefont {F.}~\bibnamefont {Schapper}}, \bibinfo {author}
  {\bibfnamefont {M.}~\bibnamefont {Holler}}, \bibinfo {author} {\bibfnamefont
  {T.}~\bibnamefont {S\"{u}dmeyer}}, \bibinfo {author} {\bibfnamefont {J.~S.}\
  \bibnamefont {Robinson}}, \bibinfo {author} {\bibfnamefont {J.~W.~G.}\
  \bibnamefont {Tisch}}, \bibinfo {author} {\bibfnamefont {F.}~\bibnamefont
  {Couny}}, \bibinfo {author} {\bibfnamefont {P.}~\bibnamefont {Light}},
  \bibinfo {author} {\bibfnamefont {F.}~\bibnamefont {Benabid}}, \ and\
  \bibinfo {author} {\bibfnamefont {U.}~\bibnamefont {Keller}},\ }\href@noop {}
  {\bibfield  {journal} {\bibinfo  {journal} {Appl. Phys. B}\ }\textbf
  {\bibinfo {volume} {97}},\ \bibinfo {pages} {369} (\bibinfo {year}
  {2009})}\BibitemShut {NoStop}%
\bibitem [{\citenamefont {Joly}\ \emph {et~al.}(2011)\citenamefont {Joly},
  \citenamefont {Nold}, \citenamefont {Chang}, \citenamefont {H\"{o}lzer},
  \citenamefont {Nazarkin}, \citenamefont {Wong}, \citenamefont {Biancalana},\
  and\ \citenamefont {{P.~{St.J}.~ Russell}}}]{Joly11}%
  \BibitemOpen
  \bibfield  {author} {\bibinfo {author} {\bibfnamefont {N.~Y.}\ \bibnamefont
  {Joly}}, \bibinfo {author} {\bibfnamefont {J.}~\bibnamefont {Nold}}, \bibinfo
  {author} {\bibfnamefont {W.}~\bibnamefont {Chang}}, \bibinfo {author}
  {\bibfnamefont {P.}~\bibnamefont {H\"{o}lzer}}, \bibinfo {author}
  {\bibfnamefont {A.}~\bibnamefont {Nazarkin}}, \bibinfo {author}
  {\bibfnamefont {G.~K.~L.}\ \bibnamefont {Wong}}, \bibinfo {author}
  {\bibfnamefont {F.}~\bibnamefont {Biancalana}}, \ and\ \bibinfo {author}
  {\bibnamefont {{P.~{St.J}.~ Russell}}},\ }\href@noop {} {\bibfield  {journal}
  {\bibinfo  {journal} {Phys. Rev. Lett.}\ }\textbf {\bibinfo {volume} {106}},\
  \bibinfo {pages} {203901} (\bibinfo {year} {2011})}\BibitemShut {NoStop}%
\bibitem [{\citenamefont {Benabid}\ \emph {et~al.}(2002)\citenamefont
  {Benabid}, \citenamefont {Knight}, \citenamefont {Antonopoulos},\ and\
  \citenamefont {{P.~{St.J}.~Russell}}}]{Benabid02a}%
  \BibitemOpen
  \bibfield  {author} {\bibinfo {author} {\bibfnamefont {F.}~\bibnamefont
  {Benabid}}, \bibinfo {author} {\bibfnamefont {J.~C.}\ \bibnamefont {Knight}},
  \bibinfo {author} {\bibfnamefont {G.}~\bibnamefont {Antonopoulos}}, \ and\
  \bibinfo {author} {\bibnamefont {{P.~{St.J}.~Russell}}},\ }\href@noop {}
  {\bibfield  {journal} {\bibinfo  {journal} {Science}\ }\textbf {\bibinfo
  {volume} {298}},\ \bibinfo {pages} {399} (\bibinfo {year}
  {2002})}\BibitemShut {NoStop}%
\bibitem [{\citenamefont {Nazarkin}\ \emph {et~al.}(2010)\citenamefont
  {Nazarkin}, \citenamefont {Abdolvand}, \citenamefont {Chugreev},\ and\
  \citenamefont {{P.~{St.J}.~Russell}}}]{Nazarkin10}%
  \BibitemOpen
  \bibfield  {author} {\bibinfo {author} {\bibfnamefont {A.}~\bibnamefont
  {Nazarkin}}, \bibinfo {author} {\bibfnamefont {A.}~\bibnamefont {Abdolvand}},
  \bibinfo {author} {\bibfnamefont {A.~V.}\ \bibnamefont {Chugreev}}, \ and\
  \bibinfo {author} {\bibnamefont {{P.~{St.J}.~Russell}}},\ }\href@noop {}
  {\bibfield  {journal} {\bibinfo  {journal} {Phys. Rev. Lett.}\ }\textbf
  {\bibinfo {volume} {105}},\ \bibinfo {pages} {173902} (\bibinfo {year}
  {2010})}\BibitemShut {NoStop}%
\bibitem [{\citenamefont {Abdolvand}\ \emph {et~al.}(2009)\citenamefont
  {Abdolvand}, \citenamefont {Nazarkin}, \citenamefont {Chugreev},
  \citenamefont {Kaminski},\ and\ \citenamefont
  {{P.~{St.J}.~Russell}}}]{Abdolvand09}%
  \BibitemOpen
  \bibfield  {author} {\bibinfo {author} {\bibfnamefont {A.}~\bibnamefont
  {Abdolvand}}, \bibinfo {author} {\bibfnamefont {A.}~\bibnamefont {Nazarkin}},
  \bibinfo {author} {\bibfnamefont {A.~V.}\ \bibnamefont {Chugreev}}, \bibinfo
  {author} {\bibfnamefont {C.~F.}\ \bibnamefont {Kaminski}}, \ and\ \bibinfo
  {author} {\bibnamefont {{P.~{St.J}.~Russell}}},\ }\href@noop {} {\bibfield
  {journal} {\bibinfo  {journal} {Phys. Rev. Lett.}\ }\textbf {\bibinfo
  {volume} {103}},\ \bibinfo {pages} {183902} (\bibinfo {year}
  {2009})}\BibitemShut {NoStop}%
\bibitem [{\citenamefont {Serebryannikov}\ and\ \citenamefont
  {Zheltikov}(2007)}]{Serebryannikov07}%
  \BibitemOpen
  \bibfield  {author} {\bibinfo {author} {\bibfnamefont {E.~E.}\ \bibnamefont
  {Serebryannikov}}\ and\ \bibinfo {author} {\bibfnamefont {A.~M.}\
  \bibnamefont {Zheltikov}},\ }\href@noop {} {\bibfield  {journal} {\bibinfo
  {journal} {Phys. Rev. A}\ }\textbf {\bibinfo {volume} {76}},\ \bibinfo
  {pages} {013820} (\bibinfo {year} {2007})}\BibitemShut {NoStop}%
\bibitem [{\citenamefont {Fedotov}\ \emph {et~al.}(2007)\citenamefont
  {Fedotov}, \citenamefont {Serebryannikov},\ and\ \citenamefont
  {Zheltikov}}]{Fedotov07}%
  \BibitemOpen
  \bibfield  {author} {\bibinfo {author} {\bibfnamefont {A.~B.}\ \bibnamefont
  {Fedotov}}, \bibinfo {author} {\bibfnamefont {E.~E.}\ \bibnamefont
  {Serebryannikov}}, \ and\ \bibinfo {author} {\bibfnamefont {A.~M.}\
  \bibnamefont {Zheltikov}},\ }\href@noop {} {\bibfield  {journal} {\bibinfo
  {journal} {Phys. Rev. A}\ }\textbf {\bibinfo {volume} {76}},\ \bibinfo
  {pages} {053811} (\bibinfo {year} {2007})}\BibitemShut {NoStop}%
\bibitem [{\citenamefont {H\"{o}lzer}\ \emph
  {et~al.}(2011{\natexlab{a}})\citenamefont {H\"{o}lzer}, \citenamefont
  {Chang}, \citenamefont {Nold}, \citenamefont {Travers}, \citenamefont
  {Nazarkin}, \citenamefont {Joly},\ and\ \citenamefont
  {{P.~{St.J}.~Russell}}}]{Hoelzer11a}%
  \BibitemOpen
  \bibfield  {author} {\bibinfo {author} {\bibfnamefont {P.}~\bibnamefont
  {H\"{o}lzer}}, \bibinfo {author} {\bibfnamefont {W.}~\bibnamefont {Chang}},
  \bibinfo {author} {\bibfnamefont {J.}~\bibnamefont {Nold}}, \bibinfo {author}
  {\bibfnamefont {J.~C.}\ \bibnamefont {Travers}}, \bibinfo {author}
  {\bibfnamefont {A.}~\bibnamefont {Nazarkin}}, \bibinfo {author}
  {\bibfnamefont {N.~Y.}\ \bibnamefont {Joly}}, \ and\ \bibinfo {author}
  {\bibnamefont {{P.~{St.J}.~Russell}}},\ }in\ \href@noop {} {\emph {\bibinfo
  {booktitle} {CLEO, Optical Society of America}}}\ (\bibinfo {year} {2011})\
  p.\ \bibinfo {pages} {CMJ3}\BibitemShut {NoStop}%
\bibitem [{\citenamefont {H\"{o}lzer}\ \emph
  {et~al.}(2011{\natexlab{b}})\citenamefont {H\"{o}lzer}, \citenamefont
  {Chang}, \citenamefont {Travers}, \citenamefont {Nazarkin}, \citenamefont
  {Nold}, \citenamefont {Joly}, \citenamefont {Saleh}, \citenamefont
  {Biancalana},\ and\ \citenamefont {{P.~{St.J}.~Russell}}}]{Hoelzer11b}%
  \BibitemOpen
  \bibfield  {author} {\bibinfo {author} {\bibfnamefont {P.}~\bibnamefont
  {H\"{o}lzer}}, \bibinfo {author} {\bibfnamefont {W.}~\bibnamefont {Chang}},
  \bibinfo {author} {\bibfnamefont {J.~C.}\ \bibnamefont {Travers}}, \bibinfo
  {author} {\bibfnamefont {A.}~\bibnamefont {Nazarkin}}, \bibinfo {author}
  {\bibfnamefont {J.}~\bibnamefont {Nold}}, \bibinfo {author} {\bibfnamefont
  {N.~Y.}\ \bibnamefont {Joly}}, \bibinfo {author} {\bibfnamefont {M.~F.}\
  \bibnamefont {Saleh}}, \bibinfo {author} {\bibfnamefont {F.}~\bibnamefont
  {Biancalana}}, \ and\ \bibinfo {author} {\bibnamefont
  {{P.~{St.J}.~Russell}}},\ }\href@noop {} {\bibfield  {journal} {\bibinfo
  {journal} {submitted}\ } (\bibinfo {year} {2011}{\natexlab{b}})}\BibitemShut
  {NoStop}%
\bibitem [{\citenamefont {Nold}\ \emph {et~al.}(2010)\citenamefont {Nold},
  \citenamefont {H\"{o}lzer}, \citenamefont {Joly}, \citenamefont {Wong},
  \citenamefont {Nazarkin}, \citenamefont {Podlipensky}, \citenamefont
  {Scharrer},\ and\ \citenamefont {{P.~{St.J}.~Russell}}}]{Nold10}%
  \BibitemOpen
  \bibfield  {author} {\bibinfo {author} {\bibfnamefont {J.}~\bibnamefont
  {Nold}}, \bibinfo {author} {\bibfnamefont {P.}~\bibnamefont {H\"{o}lzer}},
  \bibinfo {author} {\bibfnamefont {N.~Y.}\ \bibnamefont {Joly}}, \bibinfo
  {author} {\bibfnamefont {G.~K.~L.}\ \bibnamefont {Wong}}, \bibinfo {author}
  {\bibfnamefont {A.}~\bibnamefont {Nazarkin}}, \bibinfo {author}
  {\bibfnamefont {A.}~\bibnamefont {Podlipensky}}, \bibinfo {author}
  {\bibfnamefont {M.}~\bibnamefont {Scharrer}}, \ and\ \bibinfo {author}
  {\bibnamefont {{P.~{St.J}.~Russell}}},\ }\href@noop {} {\bibfield  {journal}
  {\bibinfo  {journal} {Opt. Lett}\ }\textbf {\bibinfo {volume} {35}},\
  \bibinfo {pages} {2922} (\bibinfo {year} {2010})}\BibitemShut {NoStop}%
\bibitem [{\citenamefont {Geissler}\ \emph {et~al.}(1999)\citenamefont
  {Geissler}, \citenamefont {Tempea}, \citenamefont {Scrinzi}, \citenamefont
  {Schnrer}, \citenamefont {Krausz},\ and\ \citenamefont
  {Brabec}}]{Geissler99}%
  \BibitemOpen
  \bibfield  {author} {\bibinfo {author} {\bibfnamefont {M.}~\bibnamefont
  {Geissler}}, \bibinfo {author} {\bibfnamefont {G.}~\bibnamefont {Tempea}},
  \bibinfo {author} {\bibfnamefont {A.}~\bibnamefont {Scrinzi}}, \bibinfo
  {author} {\bibfnamefont {M.}~\bibnamefont {Schnrer}}, \bibinfo {author}
  {\bibfnamefont {F.}~\bibnamefont {Krausz}}, \ and\ \bibinfo {author}
  {\bibfnamefont {T.}~\bibnamefont {Brabec}},\ }\href@noop {} {\bibfield
  {journal} {\bibinfo  {journal} {Phys. Rev. Lett.}\ }\textbf {\bibinfo
  {volume} {83}},\ \bibinfo {pages} {2930} (\bibinfo {year}
  {1999})}\BibitemShut {NoStop}%
\bibitem [{\citenamefont {Marcatili}\ and\ \citenamefont
  {Schmeltzer}(1964)}]{Marcatili64}%
  \BibitemOpen
  \bibfield  {author} {\bibinfo {author} {\bibfnamefont {E.~A.~J.}\
  \bibnamefont {Marcatili}}\ and\ \bibinfo {author} {\bibfnamefont {R.~A.}\
  \bibnamefont {Schmeltzer}},\ }\href@noop {} {\bibfield  {journal} {\bibinfo
  {journal} {Bell Syst. Tech. J.}\ }\textbf {\bibinfo {volume} {43}},\ \bibinfo
  {pages} {1783} (\bibinfo {year} {1964})}\BibitemShut {NoStop}%
\bibitem [{\citenamefont {Wegener}(2005)}]{Wegener05}%
  \BibitemOpen
  \bibfield  {author} {\bibinfo {author} {\bibfnamefont {M.}~\bibnamefont
  {Wegener}},\ }\href@noop {} {\emph {\bibinfo {title} {Extreme Nonlinear
  Optics}}}\ (\bibinfo  {publisher} {Springer-Verlag},\ \bibinfo {address}
  {Berlin},\ \bibinfo {year} {2005})\BibitemShut {NoStop}%
\bibitem [{\citenamefont {Keldysh}(1965)}]{Keldysh64}%
  \BibitemOpen
  \bibfield  {author} {\bibinfo {author} {\bibfnamefont {L.~V.}\ \bibnamefont
  {Keldysh}},\ }\href@noop {} {\bibfield  {journal} {\bibinfo  {journal}
  {Soviet Physics JETP}\ }\textbf {\bibinfo {volume} {20}},\ \bibinfo {pages}
  {1307} (\bibinfo {year} {1965})}\BibitemShut {NoStop}%
\bibitem [{\citenamefont {Sprangle}\ \emph {et~al.}(2002)\citenamefont
  {Sprangle}, \citenamefont {Pe{\~{n}}ano},\ and\ \citenamefont
  {Hafizi}}]{Sprangle02}%
  \BibitemOpen
  \bibfield  {author} {\bibinfo {author} {\bibfnamefont {P.}~\bibnamefont
  {Sprangle}}, \bibinfo {author} {\bibfnamefont {J.~R.}\ \bibnamefont
  {Pe{\~{n}}ano}}, \ and\ \bibinfo {author} {\bibfnamefont {B.}~\bibnamefont
  {Hafizi}},\ }\href@noop {} {\bibfield  {journal} {\bibinfo  {journal} {Phys.
  Rev. E}\ }\textbf {\bibinfo {volume} {66}},\ \bibinfo {pages} {046418}
  (\bibinfo {year} {2002})}\BibitemShut {NoStop}%
\bibitem [{\citenamefont {Gibson}\ \emph {et~al.}(1990)\citenamefont {Gibson},
  \citenamefont {Luk},\ and\ \citenamefont {Rhodes}}]{Gibson90}%
  \BibitemOpen
  \bibfield  {author} {\bibinfo {author} {\bibfnamefont {G.}~\bibnamefont
  {Gibson}}, \bibinfo {author} {\bibfnamefont {T.~S.}\ \bibnamefont {Luk}}, \
  and\ \bibinfo {author} {\bibfnamefont {C.~K.}\ \bibnamefont {Rhodes}},\
  }\href@noop {} {\bibfield  {journal} {\bibinfo  {journal} {Phys. Rev. A}\
  }\textbf {\bibinfo {volume} {41}},\ \bibinfo {pages} {5049} (\bibinfo {year}
  {1990})}\BibitemShut {NoStop}%
\bibitem [{\citenamefont {Augst}\ \emph {et~al.}(1991)\citenamefont {Augst},
  \citenamefont {Meyerhofer}, \citenamefont {Strickland},\ and\ \citenamefont
  {Chint}}]{Augst91}%
  \BibitemOpen
  \bibfield  {author} {\bibinfo {author} {\bibfnamefont {S.}~\bibnamefont
  {Augst}}, \bibinfo {author} {\bibfnamefont {D.~D.}\ \bibnamefont
  {Meyerhofer}}, \bibinfo {author} {\bibfnamefont {D.}~\bibnamefont
  {Strickland}}, \ and\ \bibinfo {author} {\bibfnamefont {S.~L.}\ \bibnamefont
  {Chint}},\ }\href@noop {} {\bibfield  {journal} {\bibinfo  {journal} {J. Opt.
  Soc. Am. B}\ }\textbf {\bibinfo {volume} {8}},\ \bibinfo {pages} {858}
  (\bibinfo {year} {1991})}\BibitemShut {NoStop}%
\bibitem [{\citenamefont {Agrawal}(2007)}]{Agrawal07}%
  \BibitemOpen
  \bibfield  {author} {\bibinfo {author} {\bibfnamefont {G.~P.}\ \bibnamefont
  {Agrawal}},\ }\href@noop {} {\emph {\bibinfo {title} {Nonlinear Fiber
  Optics}}},\ \bibinfo {edition} {4th}\ ed.,\ San Diego, California\ (\bibinfo
  {publisher} {Academic Press},\ \bibinfo {year} {2007})\BibitemShut {NoStop}%
\bibitem [{\citenamefont {Wood}\ \emph {et~al.}(1993)\citenamefont {Wood},
  \citenamefont {Siders},\ and\ \citenamefont {Downer}}]{Wood93}%
  \BibitemOpen
  \bibfield  {author} {\bibinfo {author} {\bibfnamefont {W.~M.}\ \bibnamefont
  {Wood}}, \bibinfo {author} {\bibfnamefont {C.~W.}\ \bibnamefont {Siders}}, \
  and\ \bibinfo {author} {\bibfnamefont {M.~C.}\ \bibnamefont {Downer}},\
  }\href@noop {} {\bibfield  {journal} {\bibinfo  {journal} {IEEE Trans. Plasma
  Sci.}\ }\textbf {\bibinfo {volume} {21}},\ \bibinfo {pages} {20} (\bibinfo
  {year} {1993})}\BibitemShut {NoStop}%
\bibitem [{\citenamefont {B\"{o}rzs\"{o}nyi}\ \emph {et~al.}(2010)\citenamefont
  {B\"{o}rzs\"{o}nyi}, \citenamefont {Heiner}, \citenamefont {Kov\'{a}cs},
  \citenamefont {Kalashnikov},\ and\ \citenamefont {Osvay}}]{Borzsonyi10}%
  \BibitemOpen
  \bibfield  {author} {\bibinfo {author} {\bibfnamefont {A.}~\bibnamefont
  {B\"{o}rzs\"{o}nyi}}, \bibinfo {author} {\bibfnamefont {Z.}~\bibnamefont
  {Heiner}}, \bibinfo {author} {\bibfnamefont {A.~P.}\ \bibnamefont
  {Kov\'{a}cs}}, \bibinfo {author} {\bibfnamefont {M.~P.}\ \bibnamefont
  {Kalashnikov}}, \ and\ \bibinfo {author} {\bibfnamefont {K.}~\bibnamefont
  {Osvay}},\ }\href@noop {} {\bibfield  {journal} {\bibinfo  {journal} {Opt.
  Express}\ }\textbf {\bibinfo {volume} {18}},\ \bibinfo {pages} {25847}
  (\bibinfo {year} {2010})}\BibitemShut {NoStop}%
\bibitem [{\citenamefont {Chang}\ \emph {et~al.}(2011)\citenamefont {Chang},
  \citenamefont {Nazarkin}, \citenamefont {Travers}, \citenamefont {Nold},
  \citenamefont {H\"{o}lzer}, \citenamefont {Joly},\ and\ \citenamefont
  {{P.~{St.J}.~Russell}}}]{Chang11}%
  \BibitemOpen
  \bibfield  {author} {\bibinfo {author} {\bibfnamefont {W.}~\bibnamefont
  {Chang}}, \bibinfo {author} {\bibfnamefont {A.}~\bibnamefont {Nazarkin}},
  \bibinfo {author} {\bibfnamefont {J.~C.}\ \bibnamefont {Travers}}, \bibinfo
  {author} {\bibfnamefont {J.}~\bibnamefont {Nold}}, \bibinfo {author}
  {\bibfnamefont {P.}~\bibnamefont {H\"{o}lzer}}, \bibinfo {author}
  {\bibfnamefont {N.~Y.}\ \bibnamefont {Joly}}, \ and\ \bibinfo {author}
  {\bibnamefont {{P.~{St.J}.~Russell}}},\ }\href@noop {} {\bibfield  {journal}
  {\bibinfo  {journal} {submitted}\ } (\bibinfo {year} {2011})}\BibitemShut
  {NoStop}%
\bibitem [{not()}]{note1}%
  \BibitemOpen
  \href@noop {} {}\bibinfo {note} {The ratio between $\tilde{\sigma}'$ and
  $\tilde{\sigma}$ is the ratio between the pulse energy contributing to the plasma
  formation and the total energy of the pulse. Full details about computing
  $\tilde{\sigma}'$ will be reported elsewhere.}\BibitemShut {Stop}%
\bibitem [{\citenamefont {Lucek}\ and\ \citenamefont {Blow}(1992)}]{Lucek92}%
  \BibitemOpen
  \bibfield  {author} {\bibinfo {author} {\bibfnamefont {J.~K.}\ \bibnamefont
  {Lucek}}\ and\ \bibinfo {author} {\bibfnamefont {K.~J.}\ \bibnamefont
  {Blow}},\ }\href@noop {} {\bibfield  {journal} {\bibinfo  {journal} {Phys.
  Rev. A}\ }\textbf {\bibinfo {volume} {45}},\ \bibinfo {pages} {6666}
  (\bibinfo {year} {1992})}\BibitemShut {NoStop}%
\bibitem [{\citenamefont {Husakou}\ and\ \citenamefont
  {Herrmann}(2001)}]{Husakou01}%
  \BibitemOpen
  \bibfield  {author} {\bibinfo {author} {\bibfnamefont {A.~V.}\ \bibnamefont
  {Husakou}}\ and\ \bibinfo {author} {\bibfnamefont {J.}~\bibnamefont
  {Herrmann}},\ }\href@noop {} {\bibfield  {journal} {\bibinfo  {journal}
  {Phys. Rev. Lett.}\ }\textbf {\bibinfo {volume} {87}},\ \bibinfo {pages}
  {203901} (\bibinfo {year} {2001})}\BibitemShut {NoStop}%
\bibitem [{\citenamefont {Zheltikov}(2008)}]{Zheltikov08}%
  \BibitemOpen
  \bibfield  {author} {\bibinfo {author} {\bibfnamefont {A.~M.}\ \bibnamefont
  {Zheltikov}},\ }\href@noop {} {\bibfield  {journal} {\bibinfo  {journal} {J.
  Raman Spectrosc.}\ }\textbf {\bibinfo {volume} {39}},\ \bibinfo {pages}
  {756} (\bibinfo {year} {2008})}\BibitemShut {NoStop}%
\end{thebibliography}
%

\end{document}